\documentclass[useAMS,usenatbib,letterpaper]{mn2e}

\usepackage{graphicx}
\usepackage{natbib}
\title[MOST observations of II~Peg]{Analysis of the MOST light curve of the heavily spotted K2IV component of
the single-line spectroscopic binary II~Pegasi\thanks{based on data from the MOST satellite,
a Canadian Space Agency mission, jointly operated by Dynacon Inc.,
the University of Toronto Institute of Aerospace Studies,
and the University of British Columbia, with the assistance
of the University of Vienna.}}

\author[M. Siwak et al.]
{Michal Siwak$^1$\thanks{E-mail: siwak@astro.utoronto.ca},
Slavek M.\ Rucinski$^1$,
Jaymie M.\ Matthews$^2$,
Rainer Kuschnig$^{2,6}$
\newauthor
David B.\ Guenther$^3$,
Anthony F.\ J.\ Moffat$^4$,
Dimitar Sasselov$^5$,
Werner W.\ Weiss$^6$\\
$^1$Department of Astronomy and Astrophysics,
University of Toronto, 50 St.\ George St., Toronto,
Ontario, M5S~3H4, Canada\\
$^2$Department of Physics \& Astronomy, University of
British Columbia, 6224 Agricultural Road, \\
Vancouver, B.C., V6T~1Z1, Canada\\
$^3$Institute for Computational Astrophysics,
Department of Astronomy and Physics,
Saint Marys University, \\  Halifax, N.S., B3H~3C3, Canada\\
$^4$D\'{e}partment de Physique, Universit\'{e}
de Montr\'{e}al, C.P.6128, Succursale: Centre-Ville,
Montr\'{e}al, QC, H3C~3J7, and\\
Centre de Recherche en Astrophysique du Qu\'{e}bec,
Canada\\
$^5$Harvard-Smithsonian Center for Astrophysics,
60 Garden Street, Cambridge, MA 02138, USA\\
$^6$Institut f\"{u}r Astronomie, Universit\"{a}t Wien,
T\"{u}rkenschanzstrasse 17, A-1180 Wien, Austria\\
}

\date{Accepted -- 28 May 2010,      Received -- ;      in original form --}

\pubyear{2010}

\begin{document}
\label{firstpage}
\maketitle

\begin{abstract}
Continuous photometric observations of the visible component
of the single-line, K2IV spectroscopic binary II~Peg
carried out by the MOST satellite during
31 consecutive days in 2008 have been analyzed.
On top of spot-induced brightness modulation,
eleven flares were detected of three distinct
types characterized by different values of rise, decay and
duration times. The flares showed a preference for occurrence
at rotation phases when the most spotted hemisphere is
directed to the observer, confirming previous similar reports.
An attempt to detect a grazing primary minimum caused by
the secondary component transiting in front of the visible star
gave a negative result. The brightness variability caused by
spots has been interpreted within a cold spot model. An assumption
of differential rotation of the primary component
gave a better fit to the  light curve than a solid-body
rotation model.
\end{abstract}

\begin{keywords}
stars: individual: II~Peg, RS~CVn-type, flare, star spots, rotation.
\end{keywords}

\section{Introduction}
\label{intro}

Studies of II~Peg (HD~224085, Lalande~46867) began with the
spectroscopic analysis of \citet{s1}. He found II~Peg to be
a late-type (K2), single-line spectroscopic binary (SB1)
and determined its first orbital elements.
Well defined, regular light variations were noticed by \citet{c1}
who explained them by rotation of the
primary component of II~Peg with a cold spot on its surface.
He also observed flares and concluded that this is a BY~Dra-type
binary system.
However, subsequent photometric and spectroscopic observations
by \citet{r1} and \citet{v1} led these authors to conclude
that the star is more akin to RS~CVn-type systems, although with
an invisible less-massive component; the distinguishing features
of the BY~Dra and RS~CVn type binaries (with, respectively,
dwarf and sub-giant components) were being defined at that time.
Changes of equivalent width of the H$_{\alpha}$ line
with orbital phase analyzed by \citet{b1} confirmed the
RS~CVn classification.

Since the 1980's, II~Peg was one of the most frequently observed
RS~CVn-type stars.
A model of multiple spots was used for analysis of
the available light-curve data sets for the first time by \citet{b2}.
The most detailed and complete study of II~Peg
was presented in a series of four papers by \citet{b3,b4,b5,b6}.
In the current paper we utilize most of the
parameters derived in the first paper of this series
which was based on high-resolution spectra used to define
a high-quality radial velocity orbit.
In brief, the essential physical parameters of the primary
(visible) star were found to be:
$T_{eff} = 4,600 \pm 100$~K, $\log~g = 3.2 \pm 0.2$,
$[Fe/H] = -0.4 \pm 0.1$, $v\, \sin i = 22.6 \pm 0.5$~km/s,
$R_{1} = 3.4 \pm 0.2$ R$_{\odot}$, spectral type K2IV,
with ephemeris for conjunction (visible star behind),
$T_{conj}=2,449,582.9268(48) + 6.724333(10) \times E$,
where $E$ is an integer number of orbits.\newline
From the analysis of TiO bands and simultaneous photometric observations,
the {\it fictitious, entirely unspotted\/}
visual magnitude of the primary star was estimated
at a relatively bright level $V_u = 6.9$; we return to this matter
later in the paper as it affects the results of our spot modelling.
The orbital inclination was estimated at $60^\circ \pm 10^{\circ}$,
leading to the primary mass $M_{1}=0.8 \pm 0.1$ M$_{\odot}$ and
implying that the secondary star is probably
a main-sequence, late-type dwarf (M0--M3V)
with mass $M_{2} \approx 0.4 \pm 0.1$ M$_{\odot}$.
The presence of a white dwarf in this binary system
was previously excluded by \citet{u1} on the basis of
ultraviolet observations made by the IUE spacecraft.

\citet{b4} presented multi-epoch images of the primary component,
obtained by means of the Doppler imaging technique.
They found that the spot distribution and spot parameters
obtained from the spectral analysis are in
good accordance with those derived solely from analysis of photometric
observations.
\citet{b6} discussed the ``flip-flop'' phenomenon, i.e.\ a shift
of the maximum spot-activity to the opposite side of the stellar surface.
The authors also concluded that -- because the largest
active area tends to be located on the hemisphere facing
the secondary star -- this component may
play an important role in the magnetic phenomena in the system.

The current paper presents analysis of continuous observations of II~Peg
conducted using the MOST satellite during 31 days in September and October 2008
(Section~\ref{obs}), a circumstance which permitted us to address the following issues:
(1)~Study of frequency and orbital-phase
localization/orientation of flares in the system (Section~\ref{flares});
(2)~A search for grazing eclipses caused by the secondary (Section~\ref{ecl});
(3)~Determination of the differential rotation of the visible star as
its minute signatures are better defined for a long observing
run (Section~\ref{lc}).

\section{Observations and data reduction}
\label{obs}

The optical system of the MOST satellite consists
of a Rumak-Maksutov f/6 15~cm reflecting telescope.
The custom broad-band filter covers the spectral range of
380 -- 700~nm with effective wavelength falling close
to Johnson's {\it V} band.
The pre-launch characteristics of the mission are described
by \citet{WM2003} and the initial post-launch performance
by \citet{M2004}.

\begin{figure}
\includegraphics[width=60mm,angle=-90]{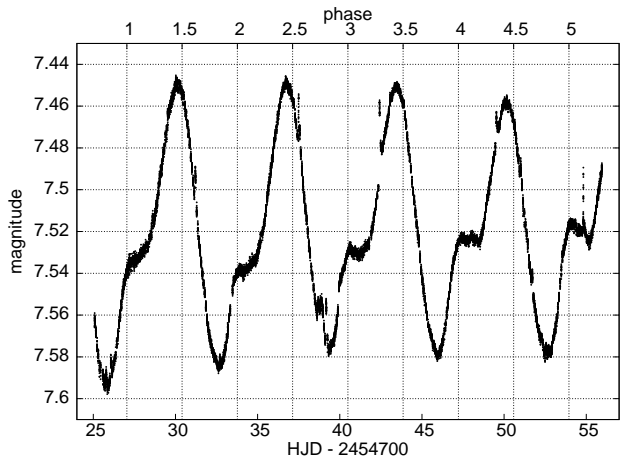}
\caption{The light curve of II~Peg in magnitudes.
The horizontal scale is in heliocentric Julian Days
(lower edge) and in orbital phase units (upper edge)
with zero phase for  conjunction with the visible star
behind. The phases  were calculated by means
of the ephemeris determined by \citet{b3} as given in
Section~\ref{intro}.
}
\label{Fig.1}
\end{figure}

II~Peg was observed from 15th September to 16th October 2008,
in $HJD = 2,454,725 - 2,454,756$,
during 439 satellite orbits over 30.877 days. The individual
exposures were 30 sec long.
Only low stray-light orbital segments were used,
lasting typically 25~min of the full 103~min satellite orbit.
In spite of the
high background, telemetry and South Atlantic Anomaly breaks,
the almost continuous light curve is very well defined
(Figure~\ref{Fig.1}).

% ----------------------- Fig.1 the light curve ---------------------
%----------------------------------------------------------------------

Because II~Peg is usually close to or slightly
fainter than 7th magnitude,
it was observed in the direct-imaging mode of the satellite
\citep{WM2003}.
The CCD camera does not have a mechanical shutter which
limits possibilities of obtaining calibration frames,
as it is commonly practised during ground-based observations.
However, \citet{r3,r4} proposed an excellent calibration procedure:
Because the background level caused by the Earth stray light
usually changes very
significantly during the orbital motion of the satellite, it
is possible to determine both the dark-level and the
flat-field information for pixels within
small images (rasters) around stars on a {\it per-pixel basis\/}.
We removed first the background gradient
visible in most frames and caused by nonuniform level of the
stray light illumination and then reconstructed the
dark and flat-field information for individual pixels on the basis
of all available frames. The final steps were
standard dark and flat-field corrections. This
approach resulted in a considerable improvement of the photometric
quality of the data. The implementation used
our own scripts written in the IDL software environment.
Aperture photometry was made by means of
DAOPHOT~II procedures \citep{stet}, as distributed by the
IDL-astro library\footnote{http://idlastro.gsfc.nasa.gov/contents.html}.

% ----------------------Fig.2 individual flares --------------------------
\begin{figure*}
\centerline{%
\begin{tabular}{l@{\hspace{3pt}}c@{\hspace{3pt}}c@{\hspace{3pt}}l}
\includegraphics[width=1.16in, angle=-90]{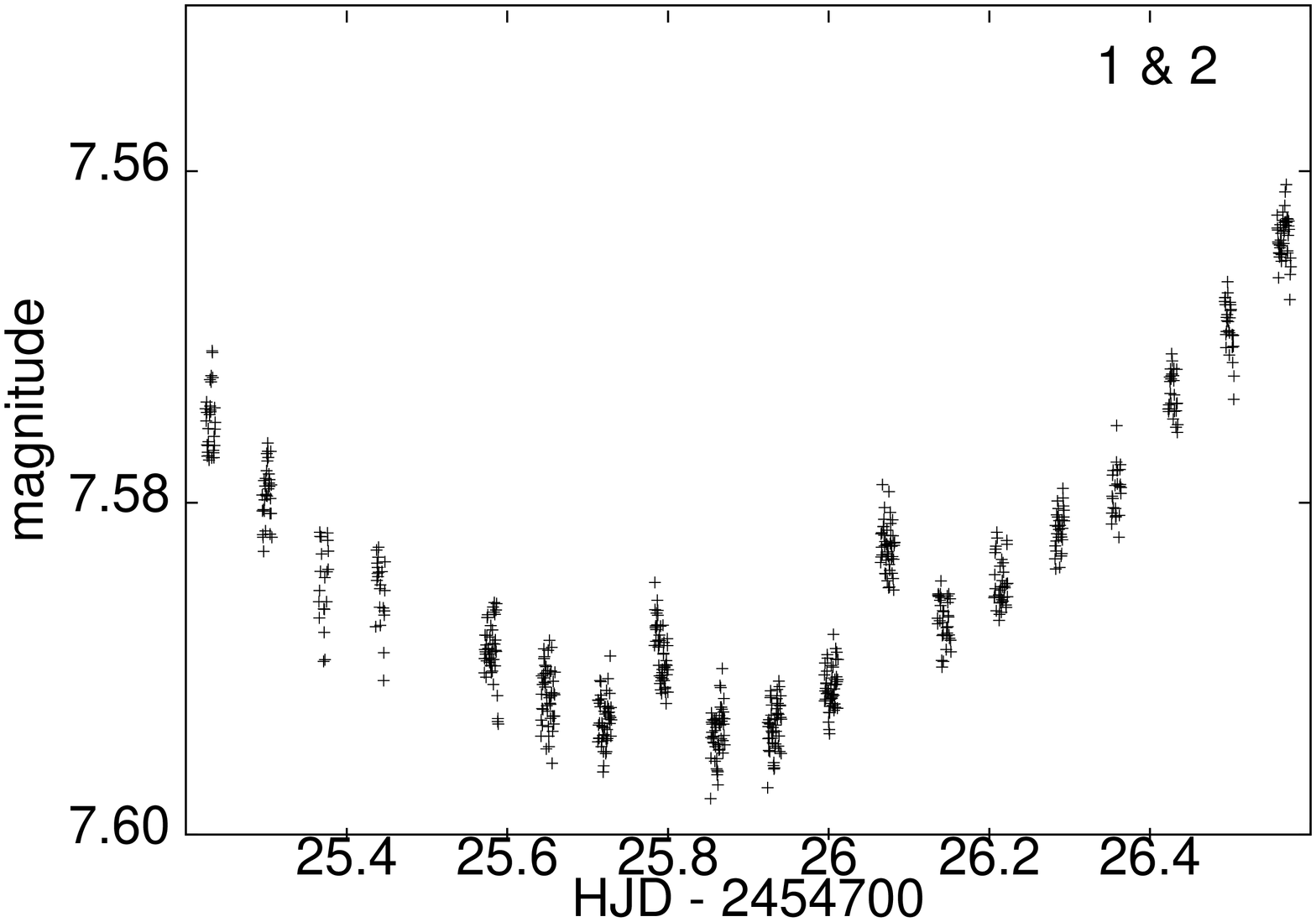} &
\includegraphics[width=1.16in, angle=-90]{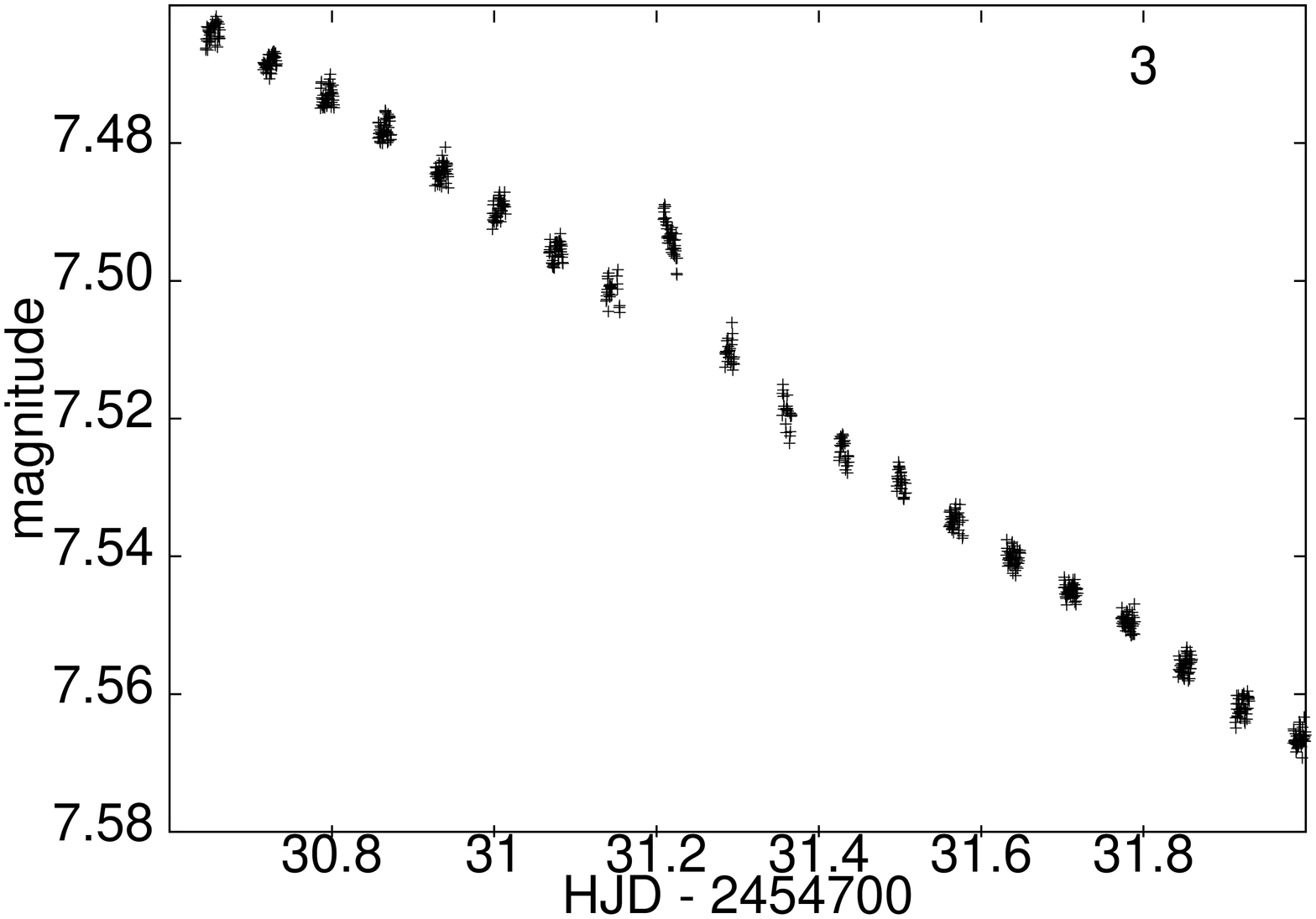} &
\includegraphics[width=1.16in, angle=-90]{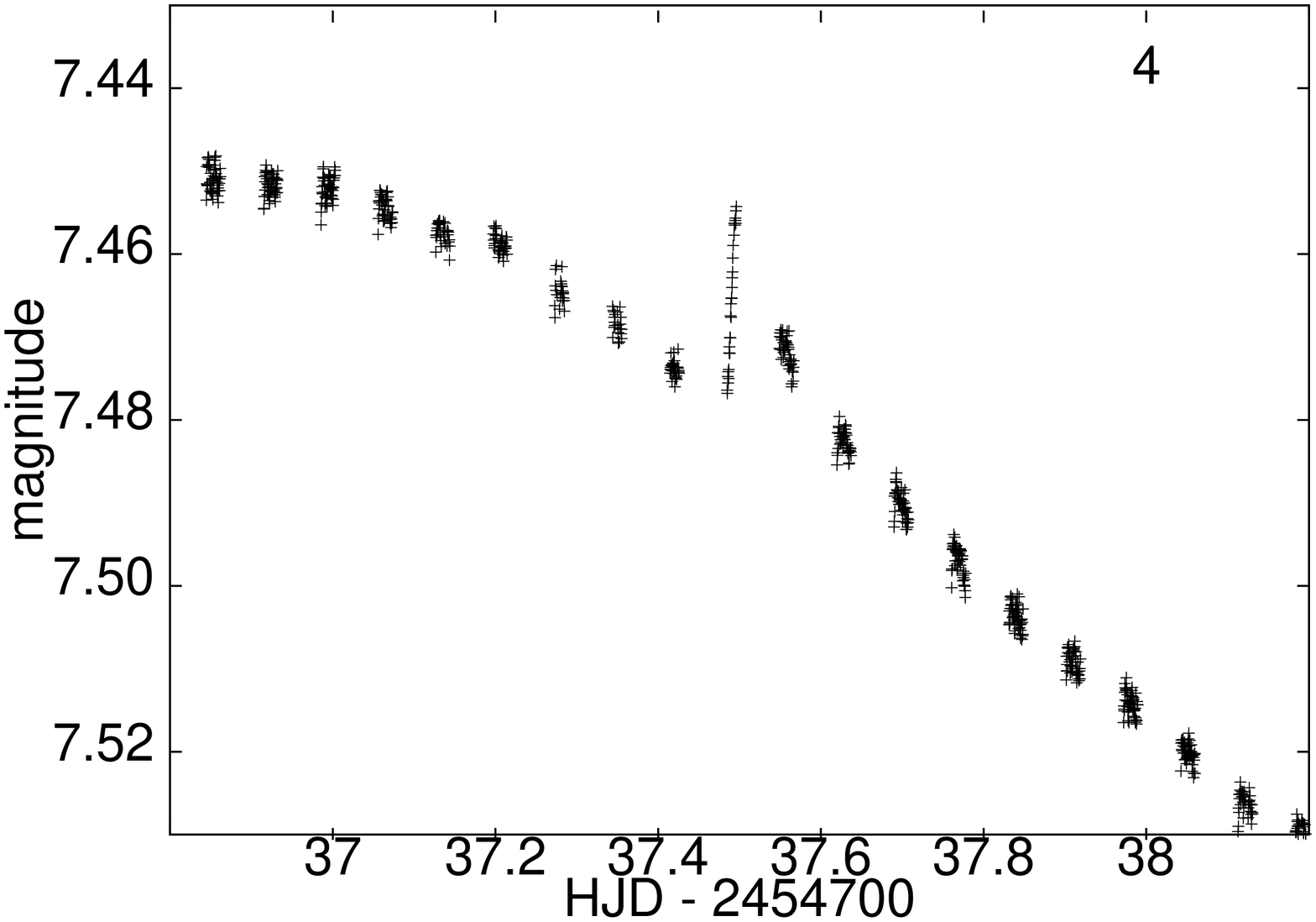} &
\includegraphics[width=1.16in, angle=-90]{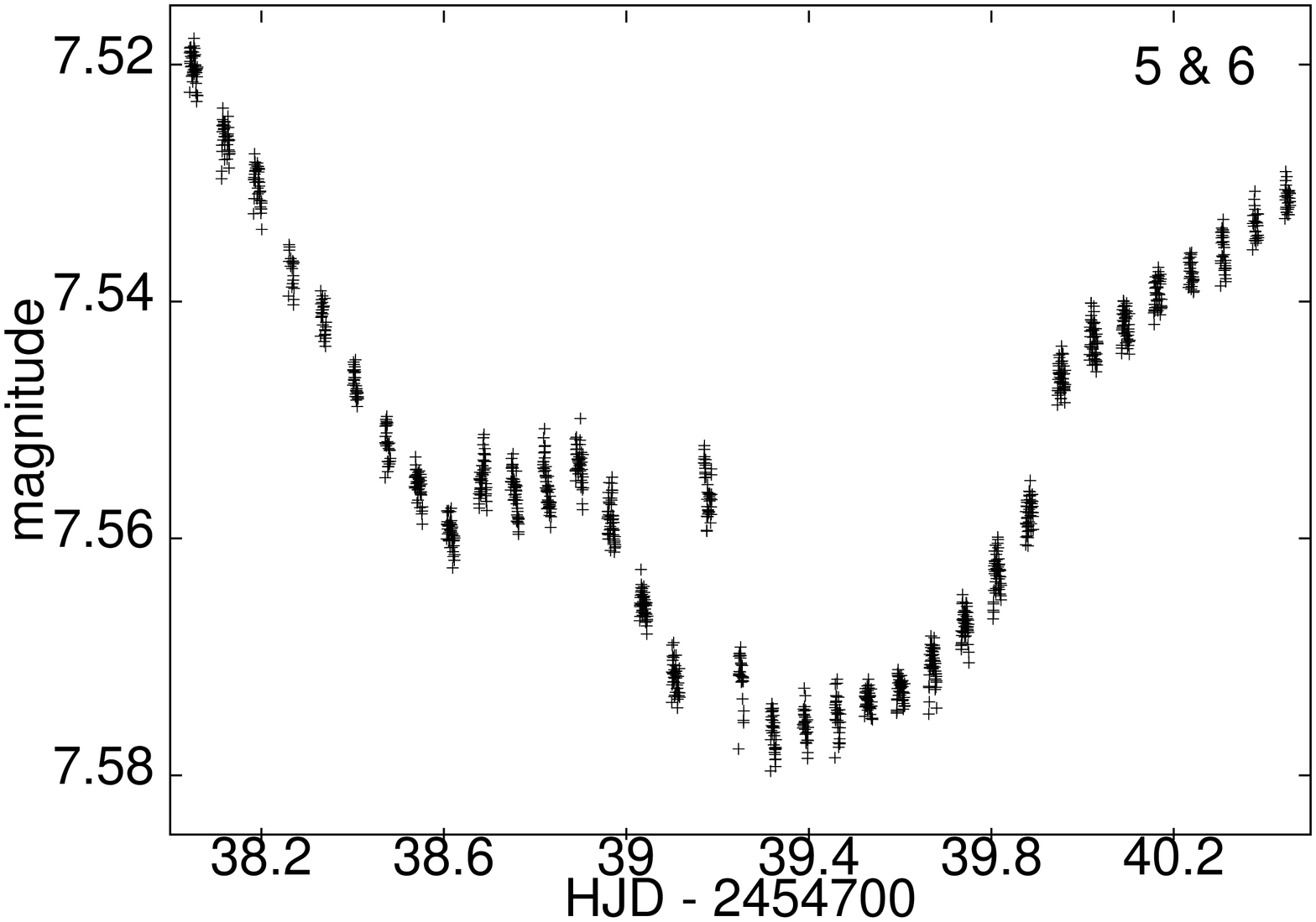} \\
\includegraphics[width=1.16in, angle=-90]{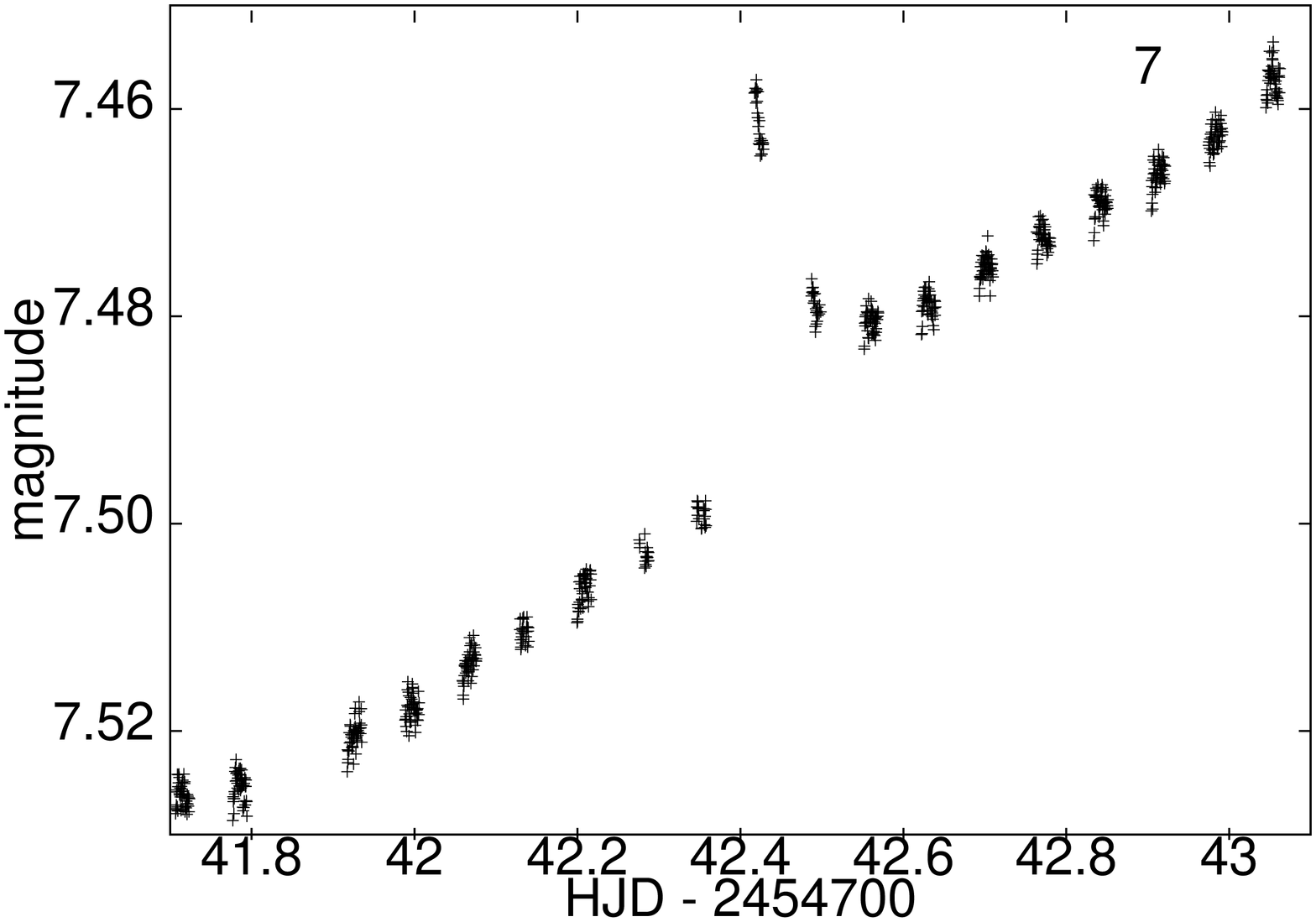} &
\includegraphics[width=1.16in, angle=-90]{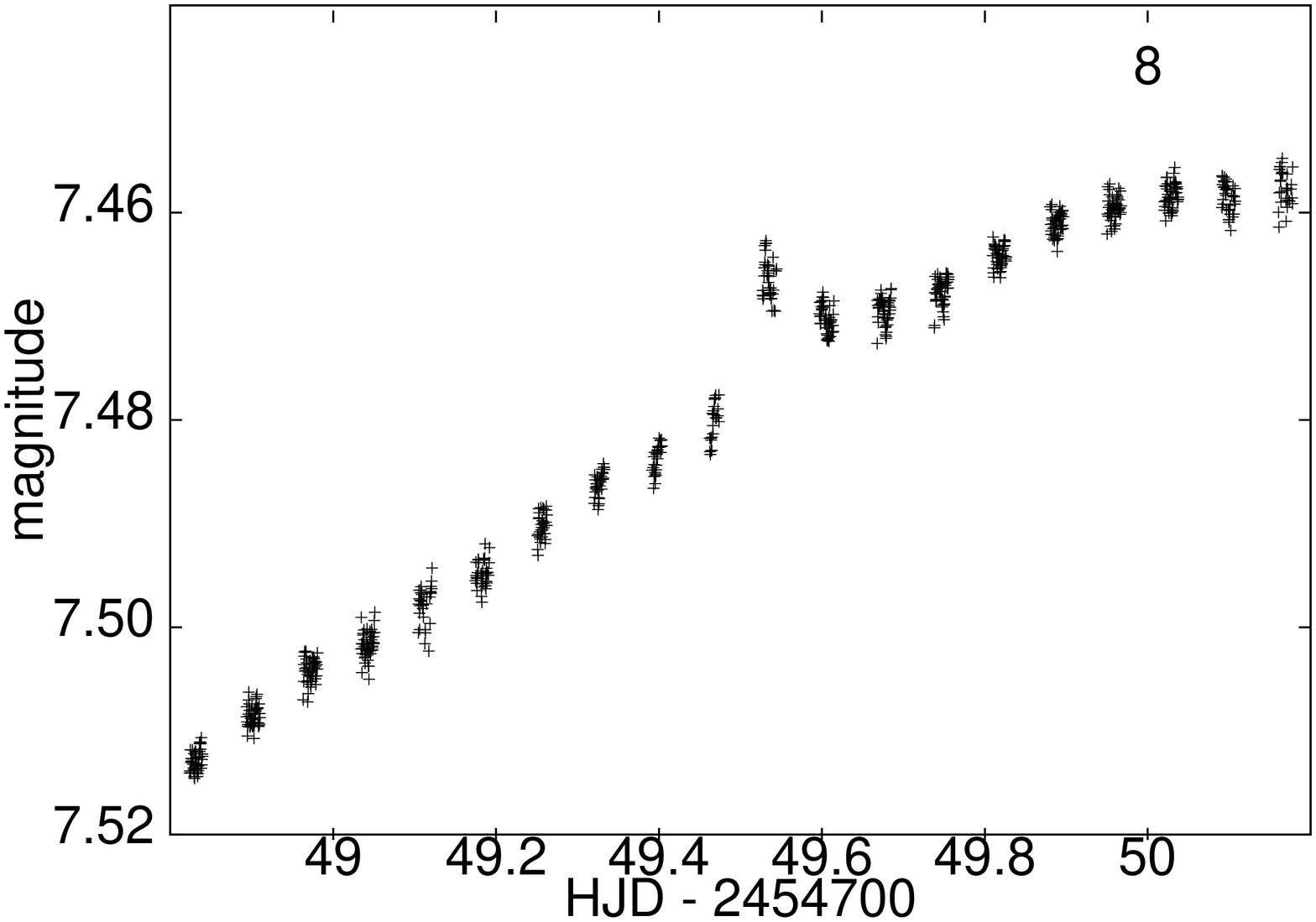} &
\includegraphics[width=1.16in, angle=-90]{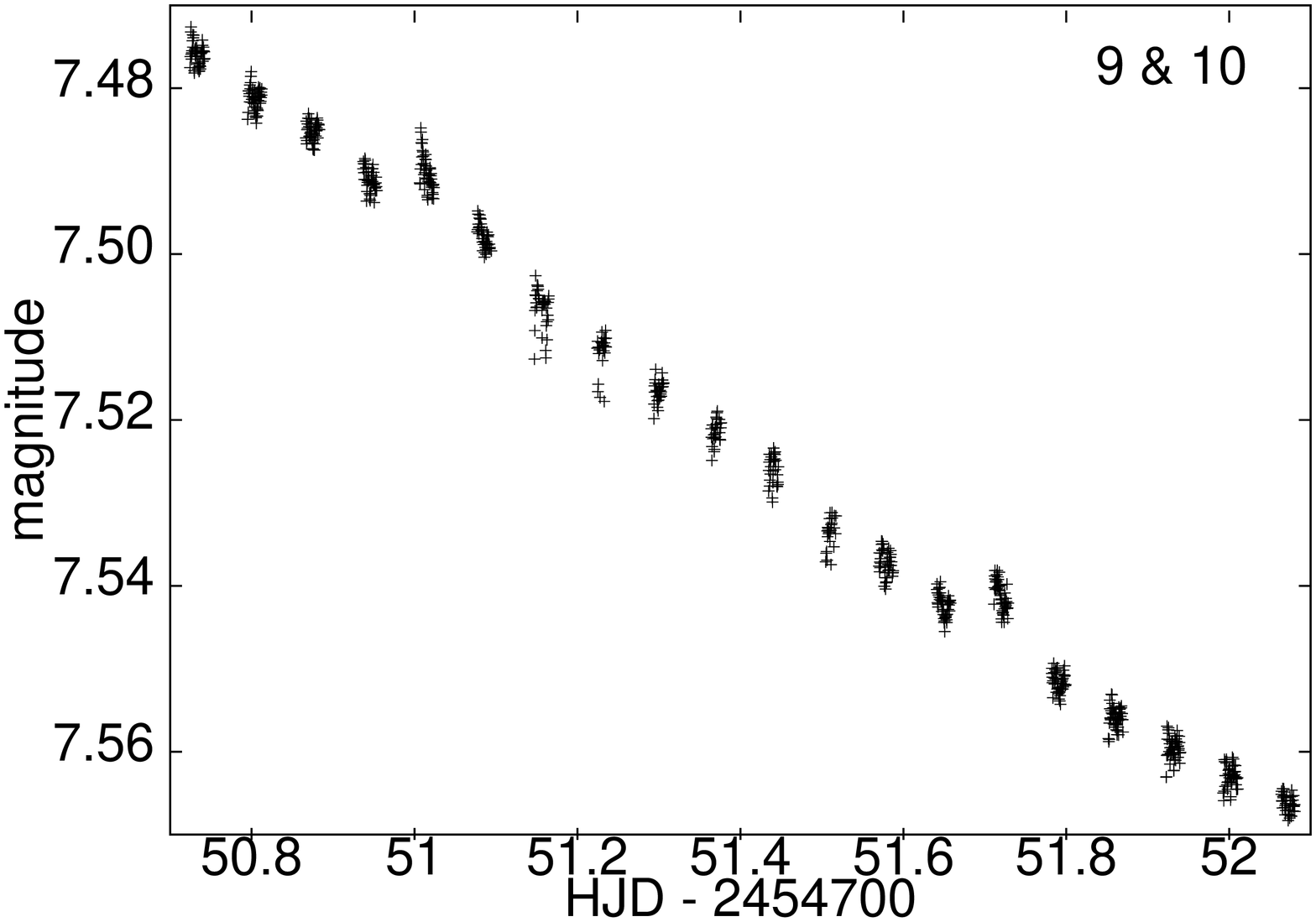} &
\includegraphics[width=1.16in, angle=-90]{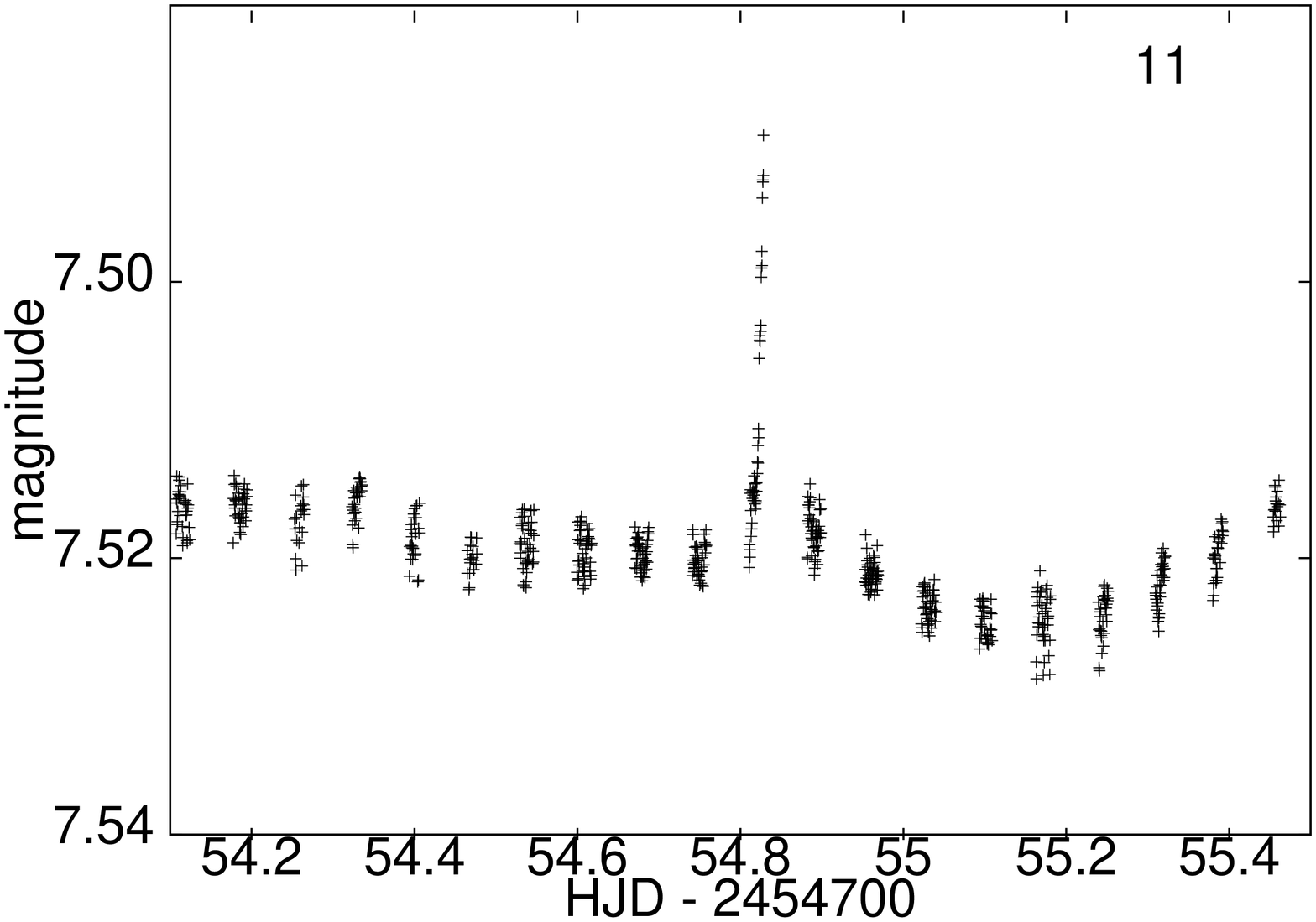}
\end{tabular}}
\caption {Enlargements of eight light-curve segments with eleven flares
of II~Peg. Groups of data points within each figure
correspond to individual {\it MOST} orbits. }
\label{Fig.2}
\end{figure*}
%----------------------------------------------------------------------------

In spite of the above careful reductions, we still observed
linear correlations between the star flux and the sky
background level, most probably caused by a small
photometric nonlinearity of the electronic system. The correlations
showed a trend with time which could be
approximated by simple linear functions of time. Corrections for the
correlations produced a smooth light curve of II~Peg with formal
scatter of about 0.002 -- 0.004 mag. However, the light curve may contain
slow (10 days or longer), smooth, systematic
trends at a level of about 0.01 magnitude which cannot be characterized and
eliminated using the available data.

The nearby, constant, simultaneously observed stars
in the unvigneted region of the CCD, GSC~02258--01385 and GSC~02258--01152,
and the low amplitude ${\delta}$~Scuti-type
star GSC~02258--00981, discovered by {\it MOST}, were used to determine transformations
between the MOST and Johnson {\it V} magnitude systems. The
$V_T$ and $B_T$ magnitudes taken from the TYCHO-2 catalogue
were used, after conversion to the standard Johnson {\it BV} system.
The maximum brightness magnitude of II~Peg during the first half of the
MOST observations was estimated at $V_{max} = 7.45 \pm 0.02$ (random)
with the additional uncertainty of the system
transformation of $\pm 0.03$ mag.
We estimate that the combined uncertainty of the maximum {\it V}-magnitude
of II~Peg during the MOST observations does not exceed $\pm 0.06$.
We note that the unspotted model prediction of \citet{b3}
was appreciably brighter, $V_u = 6.9$.

We present the light curve of II~Peg in Figure~\ref{Fig.1}, where
the {\it V} magnitudes are as determined above while
the orbital phases were calculated by means
of the ephemeris determined by \citet{b3}, as quoted in Section~\ref{intro}.
The accumulated uncertainty of the orbital period over $E= 765 - 769$ epochs
between the original determination and the MOST observations
results in a very small uncertainty of
the phase, $\pm 0.002$, which can be neglected in the present context.

We note that during the MOST observations, the upper envelope of the
light curve corresponding to the $V_{max}$ level slowly decreased from 7.45
to 7.46, while the amplitude of light changes, $\Delta V$, decreased from
0.145 to 0.12 magnitude.
It is interesting to note that the light curve obtained by {\it MOST}
is similar in its shape, maximum level and amplitude to the light curve obtained
by \citet{kal}: $V_{max}$=7.46, $\Delta V$=0.12, as presented by \citet{byr1}
and \citet{m1}.

\section{Flares}
\label{flares}

\subsection{Previous observations}
\label{hist}

The astronomical literature contains a few previous reports of several very different
flares observed for II~Peg, including cases
of non-detection even for long monitoring intervals:\newline
(1)~\citet{b1} observed sudden H$_{\alpha}$ enhancements which slowly
decayed on time scales of days;\newline
(2)~\citet{do1} simultaneously detected a flare in X-rays
and the Johnson {\it U} filter -- the latter had a duration of more than 36~min;
\newline
(3)~\citet{mat2} detected ten flares during 57.4~h of optical monitoring
in the Johnson {\it U} and {\it B} filters, finding the rate of one flare per 5.9~h;
\newline
(4a)~\citet{do2} observed two flares in their ultraviolet spectra,
with one lasting about 3~hours;\newline
(4b)~The same work reported three optical flares,
lasting 10.52, 101.00, and 9.08~min, with
amplitudes 0.066 ($B$-band), 0.371 ($U$-band) and 0.207 ($U$-band)
magnitude, respectively;\newline
(5)~\citet{m1} found one flare in their optical spectra; they summarized
the results obtained by other authors and concluded that II~Peg shows a
tendency to flare mainly when close to its minimum light;\newline
(6)~\citet{hen2} estimated a flaring rate of one flare per 4.45~h,
what agrees well with the \citet{mat2} result. They noted that \citet{byr2}
monitored II~Peg in 1992 in a $U$ filter and
found no optical flares. \citet{hen2} concluded
that II~Peg appears to exhibit long-term changes in the level of optical
flare activity;\newline
(7)~\citet{b5} observed two flares in optical
spectra, with rise times of a few hours, and very long decline times
of 1.5 and 3~d. From H$_\alpha$ emission line profiles,
they estimated that the flares had taken place above the visible
pole, probably in connection with a large, single active region; \newline
(8)~\citet{fra} found a strong flare in spectra obtained close to
light minimum, with duration time of at least 2~d.

% ---------------------------Fig.3 mean flare ----------------------------
\begin{figure}
\includegraphics[width=60mm,angle=-90]{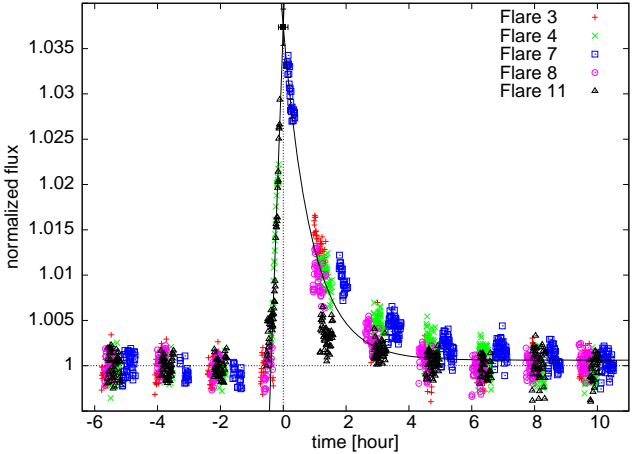}
\caption{The {\it mean flare} made from five similar,
long-lasting flares (numbers: 3, 4, 7, 8, 11),
expressed in flux units.
Time zero and maximum (1.037 continuum flux units)
correspond to the estimated moment and mean maximum amplitude of the
flare.}
\label{Fig.3}
\end{figure}
% -------------------------------------------------------------------------

%-----------------------Fig.4 brightening event-------------------

\begin{figure}
\includegraphics[width=60mm,angle=-90]{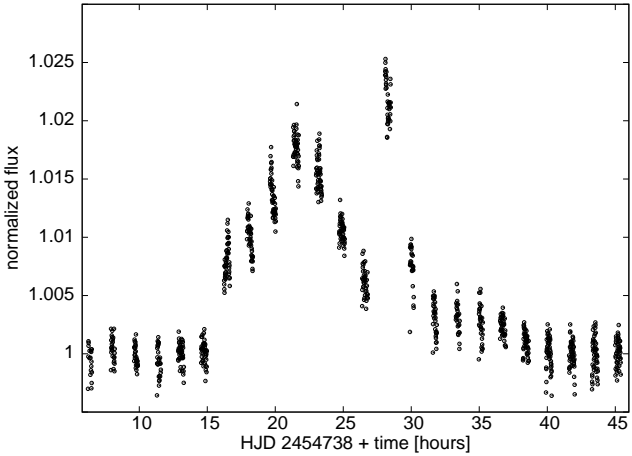}
\caption{The long-lasting flare no.5 together with the subsequent
flare no.6~.}
\label{Fig.4}
\end{figure}

%----------------------------------------------------------------

\subsection{MOST results}

Three types of flares (Fig.~\ref{Fig.2}) were detected during the MOST observations:\\
-- Four ``short'' flares (nos.\ 1, 2, 9, 10)
lasting about one or two MOST orbits
(2 -- 4~h), and with an amplitude of about 0.01~mag;\\
-- Six ``long'' flares, very similar in shape, rise and decay time,
and with amplitudes of about 0.04~mag; these flares
were used to form a ``mean flare'' (see below); \\
-- One particularly long-lasting flare, with duration of about one
full day.\\
Because of the high-background data gaps at 103~min
intervals and the typical {\it MOST} orbit coverage of 25~min,
none of the ten flares of the first two types
was observed from start to end.
In particular, the four short flares could not be analyzed
sufficiently thoroughly.

We attempted to construct a ``mean flare'' (in flux units) from the six,
apparently more commonly occurring, long-duration flares, as shown
in Figure~\ref{Fig.3}.
Because flare no.6 is affected by the preceding unusual flare no.5,
we used the remaining five flares (i.e.\ nos.\ 3, 4, 7, 8, 11) for
the construction.
First, we expressed their intensities in contiunuum flux units (as defined
by the underlying slow light variations caused by rotation of the spotted
star, removed by dividing the data by low-order polynomials fitted
to the quiescent parts of the light curve) and then we matched
the individual flare start times manually to an uncertainty
of about $\pm 2$~min. Then all flares were simply plotted
together without any further scaling;
the partially observed flares nos.\ 3 and 8
contributed only the decaying parts to such a mean flare.

The estimated rise time
from the flat continuum to the maximum of the mean flare
was found to be $25 \pm 4$~min.
The decline time, from the maximum back to the flat continuum,
varied in the range of 5 to 10~h.
The half-maximum duration time of the mean flare was
one hour ($T_{0.5} \approx 59$~min, as defined in \citet{kunk})
spanning the range between 50~min
for flare no.8 and 74~min for flare no.7. The duration
time was shorter for flare no.11,  but it could
not be uniquely determined from the available data.
All these flares considered here
most probably did not occur on the secondary component
of II~Peg, the M-dwarf, for which values of $T_{0.5}$ of the
order of hundreds of seconds would be expected \citep{kunk}; the
location was most likely the primary component or somewhere in the space
between the stars.

In general, the first two types of flares observed by {\it MOST},
were similar to those observed before by \citet{do2} (items (4a) and
(4b) in the previous section).
The long-lasting flare (no.5), which
started at $HJD=2,454,738.6$, close to light minimum,
may be an analogue of flares observed to date as enhancements
of optical spectral lines by \citet{b1}, \citet{b5} and \citet{fra}.
It differs markedly in its shape, rise time (6~h)
and decay time (at least 18~h, possibly 24~h) from the remaining
ten flares observed by {\it MOST}.
It is shown among the other flares in Figure~\ref{Fig.2}
and magnified in Figure~\ref{Fig.4}.

The rate of eleven flares in the time span of 30.877~d
gives a flaring rate of about one flare per 2.8~d.
However, due to the breaks in the MOST observations, we cannot
neglect the possibility of overlooking very short flares lasting
only a few minutes; these would be flares similar
to the two shortest observed by \citet{do2} and all of those
observed by \citet{mat2}.

%----------------------Fig.5 flares versus phase -------------------
\begin{figure}
\includegraphics[width=60mm,angle=-90]{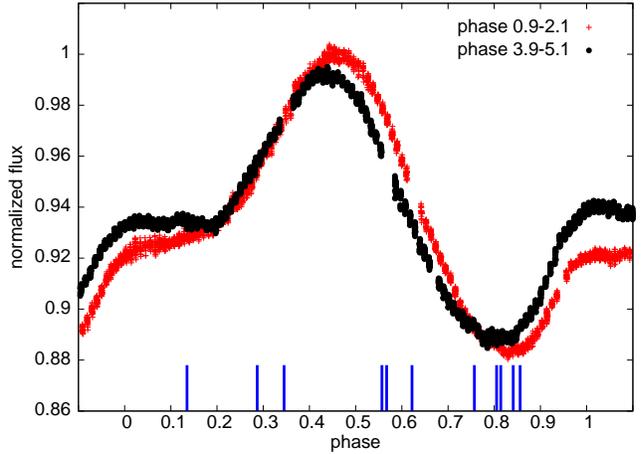}
\caption{Individual flares are indicated by vertical marks
along the horizontal (orbital phase) axis. Two sections of the
light curve of II~Peg, from the beginning (orbital
phases 0.9 -- 2.1) and the end (phases 3.9 -- 5.1) of the MOST
monitoring are shown for comparison.
The light curve evolved mainly due to a drift in
longitude of the spots at a rate faster than the orbital motion.}
\label{Fig.5}
\end{figure}
%----------------------------------------------------------------

\subsection{The phase distribution of flares}

The phase distribution of flares in relation to the
spot-modulated light curve can be inspected in detail in
Figure~\ref{Fig.5}. The flares appear not to be
uniformly distributed in orbital
phase: as many as five flares appeared  within
the light minimum, in the orbital phase interval 0.75 -- 0.85.
This supports the conclusion of \citet{m1} that flares
in II~Peg are concentrated close to light minimum
when the most heavily spotted side of the visible star is directed
toward the observer. However,
a Kolmogorov--Smirnov test for the deviation of the phase distribution
from uniformity gave the probability that the distributions appear to be
identical of 0.28. While this is a small number, it is not small
enough to prove this assertion, which still requires confirmation.

\section{A search for eclipses}
\label{ecl}

\begin{figure}
\includegraphics[width=60mm, angle=-90]{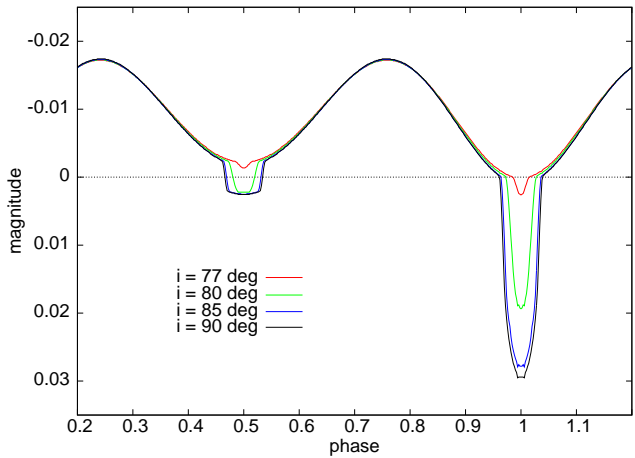}
\caption{Synthetic light curves for II~Peg computed using the stellar
parameters obtained by \citet{b3} for four different values of
the orbital inclination.}
\label{Fig.6}
\end{figure}

\begin{figure}
\includegraphics[width=60mm,angle=-90]{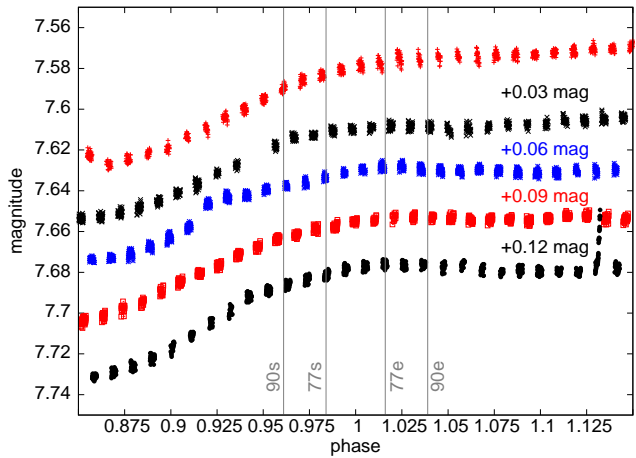}
\caption{Segments of the MOST data close to predicted times of primary eclipse.
The segments for four consecutive conjunctions (which show spot evolution
in time) have been shifted down for clarity by the indicated magnitude amounts;
the first conjunction is on top.
The start (90s, 77s) and end (77e, 90e) phases of the eclipse,
estimated for $i=90^{\circ}$ and $i=77^{\circ}$ are represented
by four vertical lines.
Note that, as discussed in Sec.~\ref{obs}, the current spectroscopic ephemeris
is very accurate and predicts the conjunction phase to $\pm 0.002$.}
\label{Fig.7}
\end{figure}

High-precision photometry from space carries a potential
for detection of eclipses caused by transits
of the undetected secondary companion over the visible star.
We estimated the expected depths and durations of primary eclipse
in the Johnson {\it V}-band for several values of inclination
using the Wilson-Devinney light curve synthetic code
\citep{wd} for the physical parameters obtained by \citet{b3};
this is shown in Fig.~\ref{Fig.6} and Fig.~\ref{Fig.7}

A careful inspection of the MOST data revealed no indication
of any eclipses, as can be seen in Figure~\ref{Fig.7}.
To analyze the conjunction segments of the light curve,
trends introduced by spots were fitted
within phase ranges 0.93 -- 0.97 and 1.03 -- 1.07
and then normalized light curves were analyzed in great detail.
The data do not reveal any systematic, localized deviations
which would have depths similar to those predicted in Fig.~\ref{Fig.6}
to less than 0.1~per~cent.

% ------------------------- Fig.6 predicted eclipses --------------------------
%-------------------------------------------------------------------------------

% -------------------------- Fig.7 MOST data for conjunctions ------------------

% ------------------------------------------------------------------------------

\begin{table*}
\centering
 \begin{minipage}{170mm}
 \caption{Results of the light curve models
 for a rigidly and a differentially rotating star for $i=60^{\circ}$.
 The ranges in resulting parameters for the two extreme
 values of the normalization parameter $F_u$, 1.19 and 1.33 (see the text)
 are given in brackets as estimates of parameter uncertainties.
 \newline a -- assumed as constant during modelling,
 \newline b -- determined using the constraints $R_1=3.4$ R$_\odot$ and $v \sin i=22.6$~km/s,
 \newline c -- calculated using Eq.(1).}
 \begin{tabular} {@{}lllll@{}}
 \hline\hline
 Prox. effects:     & neglected (1)      & neglected (2)           & accounted (3)                & accounted (4)      \\ \hline\hline
 $F_u$        & $1.26\pm0.07^a$        & $1.26\pm0.07^a$         & $1.26\pm0.07^a$            & $1.26\pm0.07^a$        \\
\hline
 $k$        & ---                      & 0.022~(0.033-0.021)$^b$& ---                          & 0.0245~(0.04-0.0225)$^b$ \\
 $P_{eq}~[d]$   & ---                      & $6.5940\pm0.0005^b$    & ---                          & $6.5940\pm0.0005^b$   \\ \hline
 $p_1~[d]$      & 6.6641 (6.6651 -- 6.6635)& 6.6748 (6.6684 -- 6.6776)$^c$ & 6.6733 (6.6738 -- 6.6729) & 6.6850 (6.6808 -- 6.6867)$^c$           \\
 $t_1~[hjd]$  & 25.903 (25.900 -- 25.903)& 25.534 (25.661 -- 25.497) & 25.824 (25.839 -- 25.821) & 25.535 (25.665 -- 25.499)\\
 ${\phi}_1~[^\circ]$ & 74.2~~~~(67.1 -- 78.3)    & 47.9~~~~(36.8 -- 50.5)       & 71.9~~~~(66.0 -- 76.6)          & 48.2~~~~(34.7 -- 51.7)    \\
 $r_1~[^\circ]$      & 31.4~~~~(24.8 -- 39.9)    & 20.5~~~~(17.9 -- 21.7)       & 31.1~~~~(25.8 -- 39.5)          & 22.4~~~~(18.8 -- 24.4)    \\ \hline
 $p_2~[d]$      & 6.6641 (6.6651 -- 6.6635)& 6.7331 (6.7432 -- 6.7309)$^c$   & 6.6733 (6.6738 -- 6.6729) & 6.7429 (6.7484 -- 6.7374)$^c$           \\
 $t_2~[hjd]$      & 28.589 (28.509 -- 28.641)& 28.018 (27.912 -- 28.057) & 28.596  (28.539 -- 28.645)   & 28.166 (28.045 -- 28.220)\\
 ${\phi}_2~[^\circ]$ & 18.1~~~~(13.6 -- 21.3)    & 75.7~~~~(55.0 -- 79.7)       & 21.2~~~~(17.7 -- 24.4)       & 71.7~~~~(49.1 -- 76.6)    \\
 $r_2~[^\circ]$      & 15.0~~~~(14.6 -- 15.4)    & 27.4~~~~(15.5 -- 36.3)       & 16.5~~~~(16.0 -- 17.2)       & 26.4~~~~(15.6 -- 34.9)    \\ \hline
 $p_3~[d]$      &  ~---                     & ~---                       & ~---                          & ~---                      \\
 $t_3~[hjd]$      &  ~---                     & ~---                       & ~---                          & ~---                      \\
 ${\phi}_3~[^\circ]$ & -90$^a$                  & -90$^a$                   & -90$^a$                      & -90$^a$                  \\
 $r_3~[^\circ]$      & ~89$^a$                  & ~89$^a$                   & ~89$^a$                      & ~89$^a$                   \\ \hline
${\chi}^2_{red,weigh}$ & 13.79         & 11.75                     & 14.72                        & 12.28             \\ \hline\hline
\end{tabular}
\label{Tab.1}
\end{minipage}
\end{table*}

\section{Light curve analysis}
\label{lc}

In this investigation we modeled the MOST light curve in terms
of dark spots which are internally invariable in time. We used 
the program {\it StarSpotz},
which was successfully applied to ${\epsilon}~Eri$ \citep{cr1} and
${\kappa^1}~Cet$ \citep{w2}, where differential rotation of the stars
was found. The reader is directed to these papers for details of the
model. The program is based on the program {\it SpotModel} \citep{rib}
which utilizes the analytical models developed by \citet{bud} and \citet{dor}.

The assumption of the internal invariability of spots is 
most likely not fulfilled in the case of II~Peg. 
As Doppler images obtained during 1994--2002
reveal \citep{b4,b6}, the spots may constantly change their properties
over time, in time scales of several rotation periods.
The shortest time scales for appreciable spot changes
could be as short as 2 months, which is comparable with the 
length of the MOST run of 31~days. 
The global, progressive light curve shape changes (Fig.~\ref{Fig.1}) 
can be easily explained by differential rotation of the stellar surface
with small, random changes of the spots well averaged over the time
of observations.  \newline
A more detailed investigation of differential rotation should be 
supported by simultaneous high-resolution spectroscopic observations.
Without them, as in the current investigation, we are unable to assert 
whether spots really remained sufficiently constant
during the {\it MOST} observations.
Additionally, whatever method of spot shape restoration is used
(including the maximum entropy method), the results will always be 
subject to limitations imposed by the instrumental effects 
mentioned in Section~\ref{obs}.

\subsection{Light curve modelling}

After first trial runs it turned out that at least two cold
spots on the hemisphere directed to the observer are necessary
to give a reasonable explanation of the light variations. 
\newline
Compatible with the non-detection of eclipses,
we assumed: $i=60^{\circ}$, $R_1=3.4$ R$_\odot$ and $v \sin i=22.6$~km/s,
as determined by \citet{b3} (see also fig. 6 of their paper).
Also, as fixed parameters for the light curve models,
we assumed the linear limb-darkening coefficient $u=0.817$,
adopted using the tables of \citet{dc1}.
We also fixed the photospheric and spot temperatures
at $T_{phot} = 4,600$~K and $T_{spot} = 3,600$~K,
as based on the results of \citet{b4,b6}.
This assumption was required to fix the spot-to-photosphere
flux ratio $f$ for the bandpass of the MOST observations.
$f = 0.077 \pm 0.015$ has been evaluated by means of
the SPECTRUM programme \citep{gray} and Kurucz's atmosphere
models \citep{kurucz} using the MOST filter bandpass.
The same value of $f$ was assumed for all spots.
The remaining parameters of the model were, for each spot:
(1) the initial moment $t$ (in $hjd \equiv HJD-2,454,700$), 
    when the spot is exactly facing the observer,
(2) the rotation period of the spot $p$ in days,
(3) the latitude ${\phi}$,
(4) the diameter $r$ in degrees and 
(5) the value of unspotted flux $F_u = 1.26$.
\newline
The latter value was calculatted assuming the value of unspotted magnitude equal to
$V_u = 7.20$, as determined by \citet{c1} at the time when
he observed a flat maximum, and also adopted by
\citet{m1}. We note that \citet{b4} suggested $V_u = 6.9$, but for
such a high brightness it is impossible to obtain
a physically plausible fit to the light curve: We would have
to postulate that we observed II~Peg with almost the whole surface
covered by  black spots.
As we described in Section~\ref{obs}, we observed $V_{max}=7.45 \pm 0.06$,
based on the MOST instrumental system, after transformation to the {\it V}-band using nearby stars.
If the unspotted magnitude for II~Peg is $V_u = 7.20$, the unspotted flux
at the time of the MOST observations would be $F_u = 1.26 \pm 0.07$ (using
normalization of $F_u = 1$ for $V_{max}=7.45$). This leads
however to the difficulty of large radii of both spots with some overlap,
which is not admitted by the model.
We solved this problem assuming a third, very
large ($r_3=89^{\circ}$) circular spot, covering practically the whole
hemisphere directed away from the observer. This spot remained constant
and -- because of the low inclination -- only partially visible.
It represented the non-variable part of the spotted photosphere.
This assumption is strongly supported by results obtained from the Doppler 
imaging technique:
According to \citet{n2}, \citet{on1}, \citet{mar} and \citet{on2} spots
are always visible and they cover between 35 to 64 per cent of the hemisphere
projected toward the observer.
As mentioned in Section~\ref{obs}, the amplitude of light changes
observed by {\it MOST} was $\Delta V = 0.145 - 0.12$ magnitude.
This was close to the smallest value noticed to date which, according
to \citet{m1}, means that during the MOST observations 
spots covered a large fraction of the stellar surface.

The light curve used in the model utilized
mean points formed for individual {\it MOST} orbits,
after removal of all stellar flares. Typically
40 -- 70 points contributed to one {\it MOST}-orbit average point.
The formal normalized flux errors ($\sigma$) per point is about 0.0011 
(median) and the full range of 0.0007 -- 0.0022.
Please note, that as discussed in Section~\ref{obs},
the light curve may contain a roughly 10~day-long smooth trend,
a few times larger in amplitude than the formal normalized flux errors
given above.

The spot model using the {\it StarSpotz} program assumes
spherical stars. II~Peg is a binary system where some light modulation
is expected from ellipsoidal and reflection effects, so-called
proximity effects.
They are small in an absolute sense (less than 0.02~mag, see
Figure~\ref{Fig.6}) but cannot be neglected as, for the assumed $i=60^{\circ}$,
they reach almost 9 per cent (0.012~mag) of the observed spot-modulation amplitude.
Because of the uncertainty with the inclination ($\pm10^{\circ}$), all physical
parameters of II~Peg are not fully known; this
affects predictions of the proximity effects.
To assess the impact of the proximity effects on the final results,
particularly on the differential
rotation of the visible star, we used two light curves,
corrected and uncorrected for the proximity effects.
Each of the two light curves was analyzed assuming a rigidly
and a differentially rotating stellar surface. 

The individual spot rotation periods $p_{1,2}$ were 
assumed to depend on the stellar latitude $\phi_{1,2}$, the rotational 
period on the equator $P_{eq}$ and differential rotation coefficient
$k$ through:
\begin{equation}
p_{i}({\phi_i})=P_{eq}/(1 - k \sin^{2}{\phi_i}),
\end{equation}
where i=1,2.
\newline
The search procedure for $k$ consisted of two steps:
first, for the assumed $R=3.4$~R$_\odot$, the proper value of
$P_{eq} = 6.5940 \pm 0.0005$~d was found, providing the observed
$v\, \sin i = 22.6$~km/s; then -- for the value of unspotted
flux level $F_u = 1.26$ -- the differential rotation coefficients $k$,
returning the smallest value of the reduced and weighted ${\chi}^2$
was derived. The formula given by Eq.(1) corresponds to the
solar-type differential rotation law.
Due to the low quality of our fits, most probably dominated
by the inadequacies of the spot model (see Section~\ref{results}),
we did not consider other possible types of differential rotation.

%------------ Fig.8 The fit to the data-----------------------------------------
\begin{figure*}
\includegraphics[height=175mm,angle=-90]{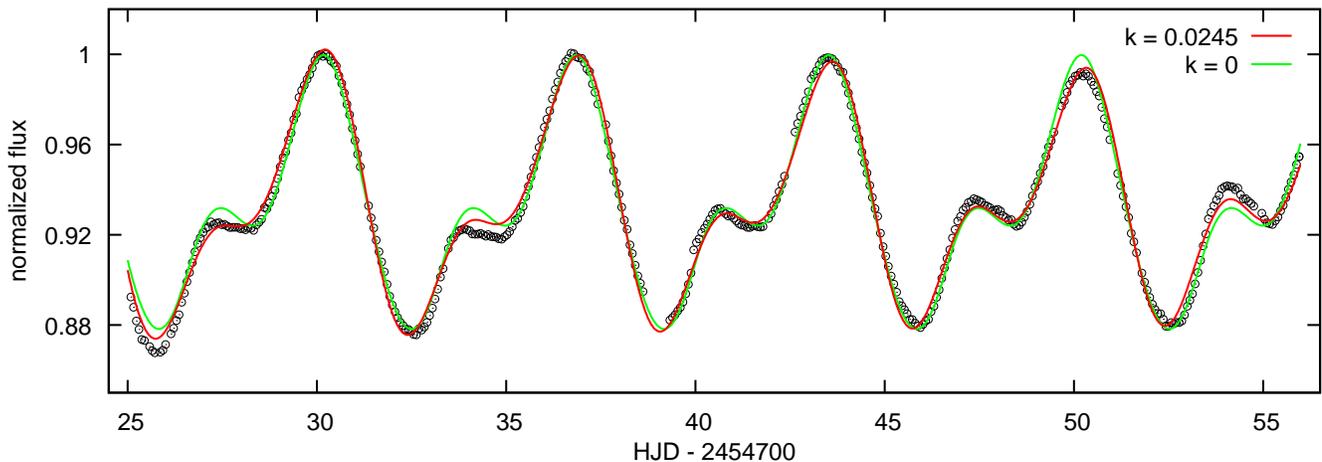}
\caption{The fit to the light curve of II~Peg (corrected for proximity effects) by the model with solid-body rotation ($k=0$, column 3 of Table~\ref{Tab.1}) 
and with differential rotation ($k=0.0245$, column 4).}
\label{Fig.8}
\end{figure*}
%-------------------------------------------------------------------------------

\subsection{Results of the light curve modelling}
\label{results}

The results of modelling are presented in Table~\ref{Tab.1}.
One can see in Figure~\ref{Fig.8}
that the fit obtained for the case of differential
rotation describes the light curve better than the solid-body rotation model.
This applies to the general light curve evolution,
and in particular to the progressive amplitude
decrease, as discussed in Section~\ref{obs}.
We also note that only for the differential-rotation models does
the larger spot face the secondary star, similarly
as obtained by \citet{b4,b6}.

To estimate the systematic errors of our models
resulting from the large uncertainty of $V_{max}$ of 
$\pm 0.06$~mag, we repeated our solutions
for two values of the unspotted flux level, $F_u = 1.19$ and
$F_u = 1.33$. This choice affects the solutions
strongly and the resulting spread in parameters can be taken
as an indication of the uncertainty in our solutions.
Note that the order in the range limits given in Table~\ref{Tab.1}
is sometimes inverted but the first value always
corresponds to the smaller value of $F_u$.

In general, the residuals -- typically at a level of
0.004 of the mean flux -- are much larger than
formal errors of individual data points (typically 0.001).
This has driven values
of the formally derived, reduced, weighted $\chi^2$ (Table~\ref{Tab.1})
to values well above unity indicating systematic trends in residuals,
most probably reflecting the difference between the true and the circular shape of spots
which was assumed in the model.\newline
Because of the dominance of the systematic deviations over the random
noise in the values of $\chi^2$, this parameter has only an indicative
utility. Nevertheless, for each pair of solutions, with
included (columns 3, 4) or excluded (columns 1, 2) proximity effects, 
the differential-rotation solution appears to be always better than the
solid-rotation one. Taking these considerations into account,
we select the solution in the last column of Table~\ref{Tab.1}
as the final one, and we plot it in Figure~\ref{Fig.8}.

\subsection{Comparison with other results}

\citet{hen} determined the differential rotation parameter,
$k=0.005 \pm 0.001$, for II~Peg
using several multi-epoch light curves.
Our result of $k = 0.0245_{-0.0020}^{+0.0155}$ is in better
accordance with the linear relation between parameters
of RS~CVn-type stars (Eq.\ 9 in \citet{hen}):
$\log k = -2.12(12) + 0.76(6) \times \log P_{rot} - 0.57(16) \times F$,
where $F=R_{star}/R_{Roche}$.
Using the parameters listed in Table~\ref{Tab.1},
we have $P_{rot} \approx 6.7$~d, $F=3.4/7.1 \approx 0.48$,
leading to a prediction of $k=0.017$.
However, when we take into account the scatter visible in
fig.~28 of \citet{hen}, a broad range of $0.002 < k < 0.066$
is admitted for this value of $P_{rot}$.
Interestingly, the value of $k$ determined in this paper
for II~Peg is similar to that estimated for the apparently single, but
even faster rotating giant, FK~Com ($k=0.016$ for $P = 2.4$~d)
by \citet{kor}.

\section{Conclusions}

Analysis of the almost-continuous, one month-long
photometric monitoring of II~Pegasi by the MOST
satellite permits us to formulate the following
conclusions:
\begin{enumerate}
\item Eleven flares were observed,
one lasting about 24~h and six flares moderately long,
lasting typically 5 to 10 hours. The characteristics of the four
shortest flares were difficult to estimate.
\item The primary eclipse
of the visible star by its companion (probably M-dwarf)
was not detected, which gives an upper limit for the orbital inclination
of the system of $76^{\circ}$.
\item From the analysis of the dark-spot modulated
light curve, assuming $i=60^{\circ}$, $R_1=3.4$~R$_\odot$,
$v \sin i=22.6$~km/s \citep{b3} and absence of internal variability
of spots during the MOST observations, we obtained
an estimate of the parameter measuring the differential rotation
of the primary component of II~Peg: $k=0.0245^{+0.0155}_{-0.0020}$.
The error of $k$ reflects the major uncertainty in the unspotted
brightness of the star so that the value of $k$
remains preliminary; it will improve with future ameliorations in
values of the assumed stellar parameters which enter the model.
\end{enumerate}

\section*{Acknowledgments}

MS acknowledges the Canadian Space Agency Post-Doctoral position
grant to SMR within the framework of the Space Science Enhancement Program.
The Natural Sciences and Engineering Research Council of
Canada supports the research of DBG, JMM, AFJM, and SMR.
Additional support for AFJM comes from FQRNT (Qu\'ebec).
RK is supported by the Canadian
Space Agency and WWW is supported by the Austrian Space
Agency and the Austrian Science Fund.

This research has made use of the SIMBAD database,
operated at CDS, Strasbourg, France and NASA's Astrophysics
Data System (ADS) Bibliographic Services.

Special thanks are due to Drs.\ Dorota Kozie{\l}-Wierzbowska
and Staszek Zo{\l}a for their attempts to detect the primary
eclipse using photometric observations of II~Peg at the
Jagiellonian University Observatory in Cracow, Poland, and to
Mr. Bryce Croll for his permission to use his spot
modelling software.


\begin{thebibliography}{41}

\bibitem[\protect\citeauthoryear{Bopp \& Noah}{1980a}]{b1}
    Bopp B.W., Noah P.V., 1980a, PASP, 92, 333

\bibitem[\protect\citeauthoryear{Bopp \& Noah}{1980b}]{b2}
    Bopp B.W., Noah P.V., 1980b, PASP, 92, 717

\bibitem[\protect\citeauthoryear{Berdyugina et al.}{1998a}]{b3}
    Berdyugina S.V., Jankov S., Ilyin I., Tuominen I., Fekel F.C.,
    1998a, A\&A, 334, 863

\bibitem[\protect\citeauthoryear{Berdyugina et al.}{1998b}]{b4}
    Berdyugina S.V.,Berdyugin A.V., Ilyin I., Tuominen I., 1998b, A\&A, 340, 437

\bibitem[\protect\citeauthoryear{Berdyugina et al.}{1999a}]{b5}
    Berdyugina S.V., Ilyin I., Tuominen I., 1999a, A\&A, 349, 863

\bibitem[\protect\citeauthoryear{Berdyugina et al.}{1999b}]{b6}
    Berdyugina S.V., Berdyugin A.V., Ilyin I., Tuominen I., 1999b, A\&A, 350, 626

\bibitem[\protect\citeauthoryear{Budding}{1977}]{bud}
    Budding E., 1977, Ap\&SS, 48, 207

\bibitem[\protect\citeauthoryear{Byrne et al.}{1989}]{byr1}
    Byrne P.B., Panagi P., Doyle J.G., Englebrecht C.A., McMahan R., Marang F., Wegner G., 1989, A\&A, 214, 227

\bibitem[\protect\citeauthoryear{Byrne et al.}{1994}]{byr2}
    Byrne P.B., Lanzafame A.C., Sarro L.M., Ryans R., 1994, MNRAS, 270, 427

\bibitem[\protect\citeauthoryear{Chugainov}{1976}]{c1}
    Chugainov P.F., 1976,
    Krymskaia Astrof.\ Obs., Izvestiia, 54, 89

\bibitem[\protect\citeauthoryear{Croll et al.}{2006}]{cr1}
Croll B., Walker G., Kuschnig R., Matthews J., Rowe J., Walker A., Rucinski S.,
Hatzes A., Cochran W., Robb R., Guenther D., Moffat A., Sasselov D., Weiss W.,
2006, ApJ, 648, 607

\bibitem[\protect\citeauthoryear{D{\'i}az-Cordov{\'e}s et al.}{1995}]{dc1}
    D{\'i}az-Cordov{\'e}s J., Claret A., Gim{\'e}nez A., 1995, A\&AS, 110, 329

\bibitem[\protect\citeauthoryear{Dorren}{1987}]{dor}
    Dorren J.D., 1987, ApJ, 320, 756

\bibitem[\protect\citeauthoryear{Doyle et al.}{1991}]{do1}
    Doyle J.G., Kellett B.J., Byrne P.B., Avgoloupis S.,
    Mavridis L.N., Seiradakis J.H., Bromage G.E., Tsuru T.,
    Makishima K., McHardy I.M., 1991, MNRAS, 248, 503

\bibitem[\protect\citeauthoryear{Doyle et al.}{1993}]{do2}
    Doyle J.G., Mathioudakis M., Murphy H.M., Avgoloupis S.,
    Mavridis L.N., Seiradakis J.H., 1993, A\&A, 278, 499

\bibitem[\protect\citeauthoryear{Frasca et al.}{2008}]{fra}
    Frasca A., Biazzo K., Tas G., Evren S., Lanzafame A.C., 2008, A\&A, 479, 557

\bibitem[\protect\citeauthoryear{Gray}{2001}]{gray}
    Gray R.O., 2001, http://phys.appstate.edu/spectrum/spe- ctrum.html,
    Department of Physics and Astronomy, Appalachian State University

\bibitem[\protect\citeauthoryear{Henry et al.}{1995}]{hen}
    Henry G.W., Eaton J.A., Hamer J., Hall, D.S., 1995, ApJSS, 97, 513

\bibitem[\protect\citeauthoryear{Henry et al.}{1996}]{hen2}
    Henry G.W., Newsom M.S., 1996, PASP, 108, 242

\bibitem[\protect\citeauthoryear{Kaluzny}{1984}]{kal}
    Kaluzny J., 1984, IBVS, No. 2627

\bibitem[\protect\citeauthoryear{Korhonen et al.}{2002}]{kor}
    Korhonen H., Berdyugina S.V., Tuominen I., 2002, A\&A, 390, 179

\bibitem[\protect\citeauthoryear{Kunkel}{1973}]{kunk}
    Kunkel W.E., 1973, ApJSS, 213,25

\bibitem[\protect\citeauthoryear{Kurucz}{1993}]{kurucz}
    Kurucz R., 1993, {\it Atomic data for opacity calculations.
    Kurucz CD-ROM No. 1.--18.}, Cambridge Mass., Smithsonian
    Astrophysical Observatory

\bibitem[\protect\citeauthoryear{Marino et al.}{1999}]{mar}
    Marino G., Rodon{\`o} M., Leto G., Cutispoto G.,
    1999, A\&A, 352, 189

\bibitem[\protect\citeauthoryear{Matthews et al.}{2004}]{M2004}
    Matthews J.M., Kusching R., Guenther D.B., Walker G.A.H.,
    Moffat A.F.J., Rucinski S.M., Sasselov D., Weiss W.W.,
    2004, Nature, 430, 51

\bibitem[\protect\citeauthoryear{Mathioudakis et al.}{1992}]{mat2}
    Mathioudakis M., Doyle J.G., Avgoloupis S., Mavridis L.N.,
    Seiradakis J.H., 1992, MNRAS, 255, 48


\bibitem[\protect\citeauthoryear{Mohin \& Raveendran}{1993}]{m1}
    Mohin S., Raveendran A.V., 1993, A\&A, 277, 155

\bibitem[\protect\citeauthoryear{O'Neal \& Neff}{1997}]{on1}
    O'Neal D., Neff J.E., 1997, AJ, 113, 1129

\bibitem[\protect\citeauthoryear{O'Neal et al.}{1998}]{on2}
    O'Neal D., Saar S.H., Neff J.E., 1998, ApJ, 501, L73

\bibitem[\protect\citeauthoryear{Neff et al.}{1995}]{n2}
    Neff J.E., O'Neal D., Saar S.H., 1995, ApJ, 452, 879

\bibitem[\protect\citeauthoryear{Udalski \& Rucinski}{1982}]{u1}
    Udalski A., Rucinski S.M., 1982, AcA, 32, 315

\bibitem[\protect\citeauthoryear{Rib{\'a}rik}{2002}]{rib}
    Rib{\'a}rik G., 2002, Occasional Technical Notes from Konkoly
    Observatory, No.12

\bibitem[\protect\citeauthoryear{Rucinski}{1977}]{r1}
    Rucinski S.M., 1977, PASP, 89, 280

\bibitem[\protect\citeauthoryear{Rowe et al.}{2006a}]{r3}
    Rowe J.F., Matthews J.M., Kusching R., et al., 2006a, Mem S.A.It., 77, 282

\bibitem[\protect\citeauthoryear{Rowe et al.}{2006b}]{r4}
    Rowe J.F., Matthews J.M., Seager S., et al., 2006b, ApJ, 646, 1241

\bibitem[\protect\citeauthoryear{Sanford}{1921}]{s1}
    Sanford R.F., 1921, ApJ, 53, 201

\bibitem[\protect\citeauthoryear{Stetson}{1987}]{stet}
    Stetson P.B., 1987, PASP, 99, 191

\bibitem[\protect\citeauthoryear{Vogt}{1979}]{v1}
    Vogt S.S., 1979, PASP, 91, 616

\bibitem[\protect\citeauthoryear{Walker et al.}{2003}]{WM2003}
Walker G., Matthews J., Kuschnig R., Johnson R., Rucinski S.,
Pazder J., Burley G., Walker A., et al., 2003, PASP, 115, 1023

\bibitem[\protect\citeauthoryear{Walker et al.}{2007}]{w2}
Walker G., Croll B., Kuschnig R., Walker A., Rucinski S., Matthews J.,
Guenther D., Moffat A., Sasselov D., Weiss W., 2007, ApJ, 659, 1611

\bibitem[\protect\citeauthoryear{Wilson}{1996}]{wd}
    Wilson R.E., 1996, {\it Documentation of Eclipsing Binary Computer Model}

\end{thebibliography}
\end{document}